\documentclass[%
reprint,
superscriptaddress,
amsmath,amssymb,
aps,
prb,
]{revtex4-1}

\usepackage{graphicx}% Include figure files
\usepackage{dcolumn}% Align table columns on decimal point
\usepackage{bm}% bold math
\usepackage{chemformula}
\usepackage{hyperref}% add hypertext capabilities
\usepackage{tabularx}% Table width
\usepackage{amsmath}% Higher math
\usepackage{float}

\newcommand{\parallelsum}{\mathbin{\!/\mkern-5mu/\!}}

\begin{document}

%\title{Phase diagram of \ch{Na_2Co_2TeO_6} single crystal under magnetic fields}
\title{Ferrimagnetism and anisotropic phase tunability by magnetic fields in \ch{Na_2Co_2TeO_6}}

\author{Weiliang~Yao}
\affiliation{International Center for Quantum Materials, School of Physics, Peking University, Beijing 100871, China}
\author{Yuan~Li}
\email[]{yuan.li@pku.edu.cn}
\affiliation{International Center for Quantum Materials, School of Physics, Peking University, Beijing 100871, China}
\affiliation{Collaborative Innovation Center of Quantum Matter, Beijing 100871, China}
\date{\today}

\begin{abstract}
\ch{Na_2Co_2TeO_6} has recently been proposed to be a Kitaev-like honeycomb magnet. To assess how close it is to realizing Kitaev quantum spin liquids, we have measured magnetization and specific heat on high-quality single crystals in magnetic fields applied along high-symmetry directions. Small training fields reveal a weak but canonical ferrimagnetic behavior below 27 K, which cannot be explained by the zigzag antiferromagnetic order alone and suggests coexisting N\'{e}el-type order of moments canted away from the zigzag chains. Moderate fields in the honeycomb plane suppress the thermal transition at 27 K, and seem to partly reverse the moment-canting when applied perpendicular to the zigzag chains. In contrast, out-of-plane fields leave the transition largely unaffected, but promotes another transition below 10 K, possibly also related to canting reversal. The magnetism in \ch{Na_2Co_2TeO_6} is highly anisotropic and close to tipping points between competing phases.

\end{abstract}

\maketitle

In the quest for quantum spin liquids (QSLs), the Kitaev honeycomb model \cite{kitaev2006,WinterJPCM2017,hermanns2018,takagi2019} is well known for its exact solvability and non-trivial properties that may be utilized in quantum computation \cite{kitaev2006,NayakRMP2008}. The model features bond-dependent Ising interactions (Kitaev interactions) between spin-1/2 degrees of freedom on a honeycomb lattice, and represents a route to magnetic frustration that is distinct from non-bipartite lattice geometries and/or competing Heisenberg interactions \cite{BalentsNature2010,ZhouRMP2017}. The lack of spin rotational symmetry, intrinsic to Kitaev interactions, appears to be at the origin of the robustness of Kitaev QSLs against perturbations of additional Heisenberg \cite{ChaloupkaPRL2010,ChaloupkaPRL2013,KatukuriNJP2014} and off-diagonal \cite{RauPRL2014,KatukuriNJP2014} exchange interactions.

Searches for Kitaev QSLs in real materials have thus been focused on realization of Kitaev interactions. It was known that spin-orbit coupling and electron correlations are essential for bond-dependent anisotropic interactions \cite{Khaliullin2005,ChenPRB2008,RauAnnRev2016}. In the pioneering work of Jackeli and Khaliullin \cite{JackeliPRL2009}, it was proposed that Kitaev interactions can be realized between spin-orbital entangled pseudospin-1/2 degrees of freedom on $d^5$ transition metal ions situated in edge-shared octahedral crystal fields, and that a honeycomb lattice of such ions may realize the Kitaev model. This proposal stimulated intense research in the past decade on 5$d$ iridium and 4$d$ ruthenium compounds \cite{WinterJPCM2017,hermanns2018,takagi2019}, with \ch{Na_2IrO_3} \cite{ChaloupkaPRL2010} and $\alpha$-\ch{RuCl_3} \cite{PlumbPRB2014} as two prominent examples. Recently, $d^7$ ions Co$^{2+}$ with a high-spin $t_{2g}^5 e_g^2$ configuration in an octahedral crystal field were proposed \cite{LiuPRB2018,SanoPRB2018} to have identical pseudospin-1/2 degrees of freedom as in the cases of $d^5$ ions Ir$^{4+}$ in \ch{Na_2IrO_3} and Ru$^{3+}$ in $\alpha$-\ch{RuCl_3}. The layered compound \ch{Na_2Co_2TeO_6} was proposed to be a candidate Kitaev material \cite{LiuPRB2018}.

To motivate the present study on \ch{Na_2Co_2TeO_6}, we first note a few commonalities and differences between this material and the earlier examples. Many of the candidate Kitaev materials known to date, including \ch{Na_2IrO_3} \cite{LiuPRB2011,ChoiPRL2012,YePRB2012} and $\alpha$-\ch{RuCl_3} \cite{SearsPRB2015,JohnsonPRB2015}, turn out to have so-called antiferromagnetic (AFM) zigzag order at low temperatures. While AFM Kitaev in conjunction with ferromagnetic (FM) Heisenberg interactions between nearest neighbors may give rise to this order \cite{ChaloupkaPRL2013}, the predicted out-of-plane moment direction is at variance with experimental observation \cite{ChunNatPhys2015}; moreover, the nearest-neighbor Kitaev interactions in both \ch{Na_2IrO_3} \cite{ChaloupkaPRL2010,KatukuriNJP2014,YamajiPRL2014,SizyukPRB2014,HuPRL2015,WinterPRB2016} and $\alpha$-\ch{RuCl_3} \cite{WinterPRB2016,KimPRB2016,RanPRL2017,BanerjeeNPJQM2018} are believed to be FM instead of AFM, so there must be additional interactions, off-diagonal and/or further-neighbor Heisenberg ones in particular, in order to stabilize the zigzag order \cite{ChunNatPhys2015,KatukuriNJP2014,RauPRL2014,SizyukPRB2014,WinterPRB2016,YamajiPRL2014,KimchiPRB2011} and explain related thermodynamic properties \cite{SinghPRL2012,DasPRB2019}. \ch{Na_2Co_2TeO_6} is no exception in these regards -- both zigzag order \cite{LefrancoisPRB2016,BeraPRB2017} and nearest-neighbor FM Kitaev interactions \cite{LiuPRB2018,SanoPRB2018} have been found. Nevertheless, it might be advantageous as a starting point to realize Kitaev QSLs, because unlike \ch{Na_2IrO_3} and $\alpha$-\ch{RuCl_3}, \ch{Na_2Co_2TeO_6} has no monoclinic distortion \cite{LefrancoisPRB2016,BeraPRB2017,ViciuJSSC2007}, and the $d^7$ high-spin configuration brings about a cancelation mechanism for the nearest-neighbor Heisenberg interactions \cite{LiuPRB2018,SanoPRB2018}. Yet still, the existence of additional interactions remains a major factor of unknowns.

Here we report comprehensive magnetization and specific heat measurements on high-quality \ch{Na_2Co_2TeO_6} single crystals. In essentially zero magnetic field, we establish ferrimagnetic behaviors in, and only in, the previously identified zigzag AFM phase. This in turn suggests an intrinsic admixture of N\'{e}el-type AFM order with the zigzag order, which also leads to non-collinear moment canting. We further show that moderate magnetic fields can suppress the thermal transition into the zigzag phase when the field is applied parallel to the honeycomb plane. Finally, while ordered moments in the zigzag phase point mainly along the zigzag chains, their canting can be partly reversed by moderate transverse fields. We believe that this kaleidoscope of phase behaviors will help restrict future explorations on the magnetic interactions and navigate the tuning of \ch{Na_2Co_2TeO_6} towards Kitaev QSLs.

\begin{figure}
	\centering{\includegraphics[clip,width=8cm]{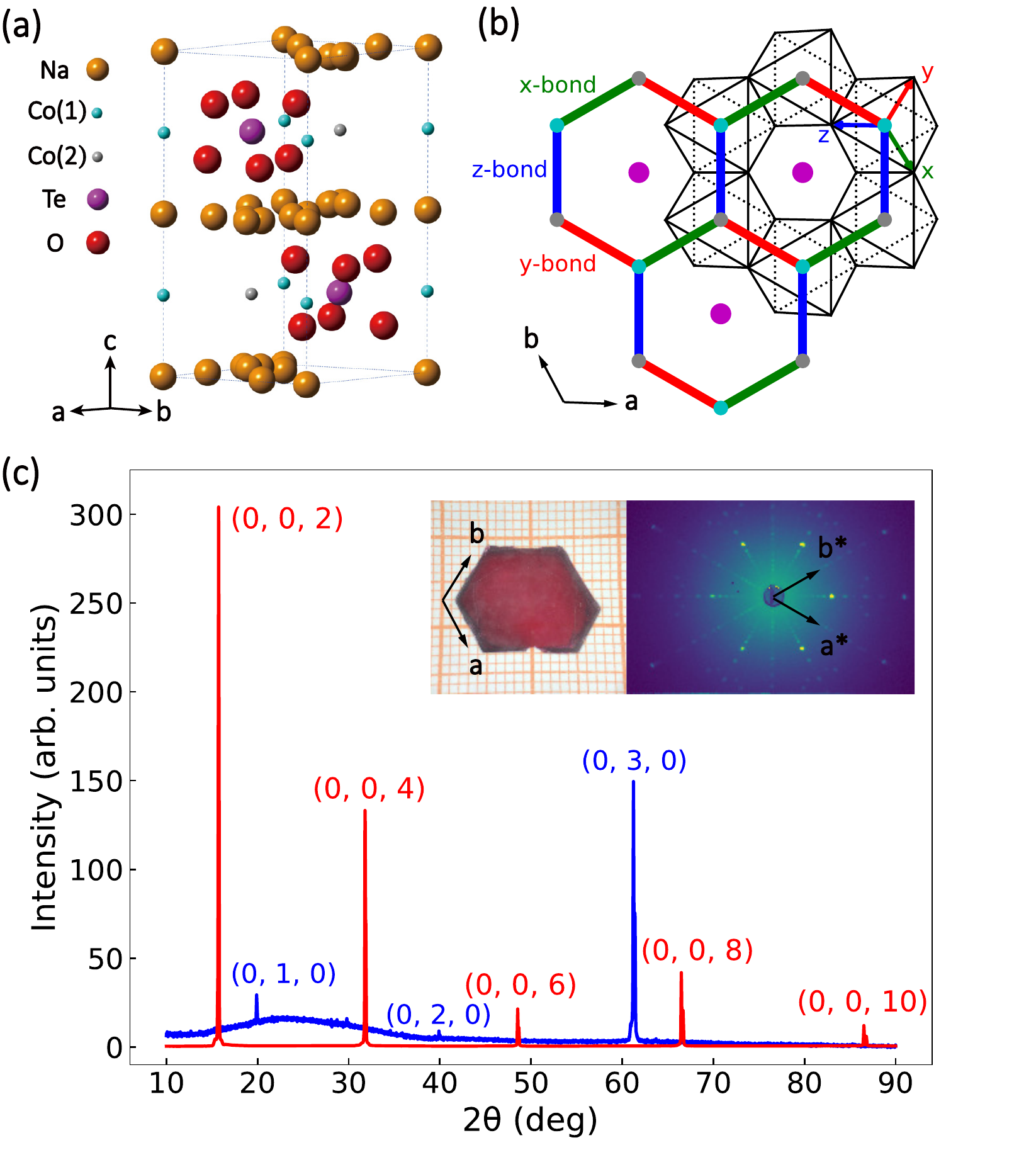}}
	\caption{(a) Crystal structure of \ch{Na_2Co_2TeO_6}. (b) The honeycomb layer viewed along $\mathbf{c}$. Idealized \ch{CoO_6} octahedra are shown in one of the hexagons. Kitaev interactions between the $x$, $y$, and $z$ components of the pseudospins are indicated by thick colored lines, with the corresponding component directions shown in one of the octahedra. %Grey rectangle indicates a \ch{Co_2O_6} double-plaquette perpendicular to the pseudospin $z$-axis.
(c) X-ray diffraction data taken on a $(0, 0, L)$ and a $(0, K, 0)$ crystal surface. The broad hump below $40^\circ$ is background. Inset shows a representative crystal on a millimeter grid (left) and the coresponding X-ray Laue pattern (right).}
	\label{fig1}
\end{figure}

The structure of \ch{Na_2Co_2TeO_6} [Fig.~\ref{fig1}(a)] belongs to space group $P6_322$ (No. 182) \cite{LefrancoisPRB2016,BeraPRB2017,ViciuJSSC2007}. It contains edge-sharing \ch{CoO_6} (and \ch{TeO_6}) octahedra that form a perfect honeycomb lattice of the \ch{Co^{2+}} ions [Fig.~\ref{fig1}(b)]. The honeycomb layers are sandwiched between \ch{Na^{+}} layers, and they stack along the $\mathbf{c}$ direction such that a two-fold screw axis goes through the $A$-sublattice [Wyckoff $2b$, Co(1)] of the \ch{Co^{2+}} honeycomb. \ch{Co^{2+}} ions on the honeycomb $B$-sublattice [Wyckoff $2d$, Co(2)] are stacked atop \ch{Te^{6+}} in adjacent honeycomb layers, and their environment due to \ch{Na^{+}} ions is somewhat different from that of Co(1). Nevertheless, the oxygen octahedra surrounding Co(1) and Co(2) are similar, with the Co-O distances differing by about 1\% \cite{BeraPRB2017,ViciuJSSC2007}. In such approximate octahedral crystal fields, the $d^7$ \ch{Co^{2+}} ions are in their high-spin $t_{2g}^{5}e_{g}^{2}$ configuration with $S=3/2$ and $L=1$. When non-cubic crystal field \cite{BeraPRB2017} is weaker than spin-orbit coupling and the neighboring Co(1)-O-Co(2) angle close to $90^\circ$ (about $92^\circ$ in this case \cite{ViciuJSSC2007,BeraPRB2017}), the resultant atomic ground state is a doublet, and Kitaev interactions exist between these pseudospin-1/2 degrees of freedom \cite{LiuPRB2018,SanoPRB2018} [Fig.~\ref{fig1}(b)]. Neutron diffraction studies \cite{LefrancoisPRB2016,BeraPRB2017} found long-range zigzag AFM order below $T_\mathrm{N} \sim$ 27 K with propagation vector $(1/2, 0, 0)$ (and its equivalent). The ordered magnetic moments on \ch{Co^{2+}} were believed to point along the zigzag chains.

The high-quality single crystals of \ch{Na_2Co_2TeO_6} used in this study were grown with a flux method modified from a recent report \cite{XiaoCGD2019}. Crystals were ruby-colored hexagonal flakes of typical size  $\sim10\times10\times0.1$ mm$^3$. Chemical stoichiometry was confirmed by energy dispersive spectroscopy (Oxford, see Fig.~S1 and Table~S1 in \cite{SM}). X-ray backscattering Laue (Photonic Science) patterns taken on large crystal faces oriented as in inset of Fig.~\ref{fig1}(c) had six-fold symmetry, indicating the face to be the honeycomb plane. Single-crystal X-ray diffraction (Rigaku, Cu K$_\alpha$) produced sharp $(0, 0, L)$ reflections with low background [Fig.~\ref{fig1}(c)], which suggested good crystallinity. Using a thick crystal, we were further able to perform diffraction on its side (Fig.~S2 in \cite{SM}), and the observed peaks could be indexed as $(0, K, 0)$ reflections [Fig.~\ref{fig1}(c)], hence allowing us to determine a crystal's orientation by its shape. All magnetic field directions used in the following measurements were based on this result.

Our DC magnetization and specific heat measurements were performed with a Quantum Design MPMS and PPMS, with fields up to 7 and 9 T, respectively. All presented data on \ch{Na_2Co_2TeO_6} were acquired on the same crystal of $\sim$1.64 mg, and our reference specific heat measurement on \ch{Na_2Zn_2TeO_6} was performed on a crystal of $\sim$1.17 mg (Fig.~S3 in \cite{SM}). To check consistency with previous studies, we measured magnetization over a wide temperature range (Fig.~S4 in \cite{SM}). A Curie-Weiss fit of data between 200 K and 300 K resulted in effective magnetic moments of 5.69 $\mu_\mathrm{B}$ and 5.31 $\mu_\mathrm{B}$ per Co$^{2+}$, and Weiss temperatures $\Theta$ of 12.5 K and -93.8 K, respectively, for $H\parallelsum\mathbf{a}^*$ and $H\parallelsum\mathbf{c}$. These values are consistent with previous reports \cite{LefrancoisPRB2016,BeraPRB2017,XiaoCGD2019}. In particular, the slightly positive $\Theta_{\mathbf{a}*}$ (less negative $\Theta_\mathbf{ab}$ than $\Theta_\mathbf{c}$ in \cite{XiaoCGD2019}) is encouraging, since it indicates the presence of ferromagnetic interactions. While further-neighbor interactions may still be part of the origin, this is consistent with the expected ferromagnetic Kitaev and weak Heisenberg interactions between the nearest neighbors \cite{LiuPRB2018,SanoPRB2018}.

\begin{figure}
	\includegraphics[clip,width=8.5cm]{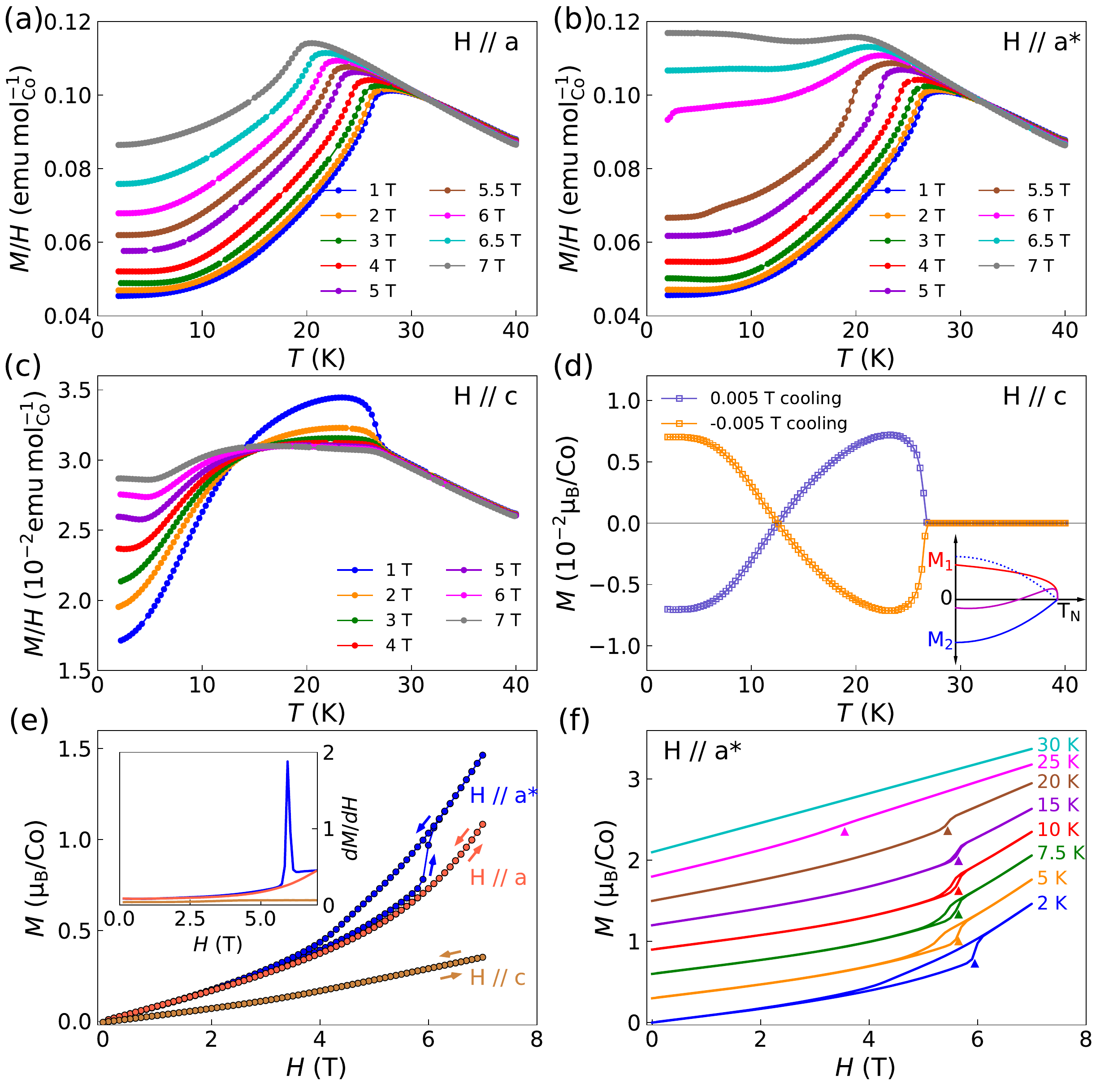}
	\caption{(a-c) Magnetization ($M$) divided by $H$ along different directions. The measurements in (a) and (b) were performed after zero-field-cooling, and in (c) after field-cooling, in order to avoid irregularities due to undesired training of the ferrimagnetism (see text). (d) Residual magnetization after cooling in weak training fields along \textbf{c}, measured with no field. Inset illustrates how ferrimagnetism gives rise to residual magnetization reversal (see text). (e) Field dependence of magnetization at 2 K. Only the $H\parallelsum\mathbf{a}^*$ measurement shows a jump (inset shows field derivatives in units of $\mu_\mathrm{B}$/T per Co$^{2+}$) along with hysteretic behavior near 6 T, and this first-order transition field decreases with increasing temperature [panel (f), offset in 0.3 $\mu_\mathrm{B}$/Co increments for clarity].}
	\label{fig2}
\end{figure}

We first present in Fig.~\ref{fig2}(a-c) variable-$T$ magnetization measurements with fields applied in three high-symmetry directions: along the zigzag chains ($H\parallelsum\mathbf{a}$), in-plane but perpendicular to the chains ($H\parallelsum\mathbf{a}^*$), and out-of-plane ($H\parallelsum\mathbf{c}$), see Fig.~\ref{fig1}. The two in-plane-field geometries should be understood with the presence of three 120$^\circ$-different zigzag domains in our sample below $T_\mathrm{N}$ -- we have tried cooling the crystal with our largest fields in both directions, but observed no magnetic ``detwinning'' effects. The results in Fig.~\ref{fig2}(a-c) offer a first glance at our first main finding: For both $H\parallelsum\mathbf{a}$ and $H\parallelsum\mathbf{a}^*$, $T_\mathrm{N}$ (as seen from the steepest variation of $M/H$ versus $T$) is continuously suppressed by the increasing fields; in contrast, $T_\mathrm{N}$ manifests itself as an anomaly in $M$ for $H\parallelsum\mathbf{c}$, and it does not change much with field. The same data reveal possible field-induced transitions for $H\parallelsum\mathbf{a}^*$ between 5.5 and 6 T as seen from the extra increase in $M/H$ below $T_\mathrm{N}$, and for $H\parallelsum\mathbf{c}$ at about 4 T, since a small upturn in $M$ appears below $\sim5$ K at fields higher than this. We will come back to these points later.

It is useful here to compare our results with some previous reports. The decrease of $T_\mathrm{N}$ and the recovery of suppressed $M/H$ below $T_\mathrm{N}$ with increasing in-plane fields have also been found in $\alpha$-\ch{RuCl_3} \cite{BanerjeeNPJQM2018,YuPRL2018}, which has been suggested to eventually lead to a field-induced QSL state \cite{BanerjeeNPJQM2018,KasaharaPRL2018,KasaharaNature2018,WellmPRB2018}. In our case, $M/H$ becomes $H$-independent starting from temperatures slightly above $T_\mathrm{N}$ (in the zero-field limit), whereas in $\alpha$-\ch{RuCl_3} a pronounced $H$ dependence remains far above $T_\mathrm{N}$ \cite{SearsPRB2017,BanerjeeNPJQM2018,YuPRL2018,ShiPRB2018}. We do not understand this difference at present -- although $T_\mathrm{N}$ is higher in our case, the effective moments are also over twice larger, so we expect the field-related energies to work against thermal fluctuations up to higher $T$ as well.

Compared to the rather complicated behaviors below $T_\mathrm{N}$ previously found in measurements on \ch{Na_2Co_2TeO_6} using smaller fields \cite{LefrancoisPRB2016,BeraPRB2017,XiaoCGD2019}, our results in Fig.~\ref{fig2}(a-c) look much simpler. Although an extra anomaly at about 16 K can indeed be observed in our crystal with smaller fields (Fig.~S5 in \cite{SM}), we uncover here an aspect that can easily mislead low-field measurements: the system is ferrimagnetic below $T_\mathrm{N}$. This is most evident when the crystal is prepared by cooling in a weak training field parallel to $\mathbf{c}$ and then measured in zero field [Fig.~\ref{fig2}(d)]. A small yet clear \textit{negative} magnetization remains after the field is turned off, and it changes sign at about 12.5 K, before vanishing above $T_\mathrm{N}$. This is a canonical ferrimagnetic phenomenon \cite{KumarPhysRep2015}, where each of two magnetic sublattices of an antiferromagnet has a net total moment, unequal and opposite to each other, and the moment that initially grows faster below $T_\mathrm{N}$ saturates at a smaller value at the lowest $T$ [inset of Fig.~\ref{fig2}(d)]. The sum therefore reverses sign at an intermediate $T$ called the compensation point \cite{KumarPhysRep2015}, 12.5 K in our case (see also Fig.~S6 in \cite{SM}). The uncompensated magnetization turns out to be greater along $\mathbf{c}$ than in-plane, making it still visible in the data in Fig.~\ref{fig2}(c), but the ferrimagnetism certainly has an in-plane part as well (Fig.~S5 in \cite{SM}), which complicates low-field measurements especially if nominal zero-field-cooling actually involves an uncontrolled remnant field.

The presence of ferrimagnetism, which likely arises from the two Co$^{2+}$ sublattices, has important implications. With no structural distortion reported below $T_\mathrm{N}$ \cite{LefrancoisPRB2016,BeraPRB2017,XiaoCGD2019}, the zigzag order does \textit{not} have net moments on either of the sublattices, and cannot be its origin. Only intra-unit-cell ($\mathbf{q}=0$) N\'{e}el order can cause such behaviors. As the ferrimagnetism is found only below $T_\mathrm{N}$, we attribute it to an admixture of N\'{e}el order with the zigzag order, the former of which involves slight moment canting away from the zigzag chains and remains hitherto undetected by neutron diffraction. The combined order is non-collinear and may be favored by off-diagonal interactions \cite{RauPRL2014,KatukuriNJP2014,ChaloupkaPRB2016,RusnackoPRB2019}. Moreover, the magnetization reversal at 12.5 K implies that distinct intra-sublattice interactions, \textit{e.g.}, between next-nearest neighbors in the honeycomb lattice, are present in the system.

\begin{figure}
	\centering{\includegraphics[clip,width=8cm]{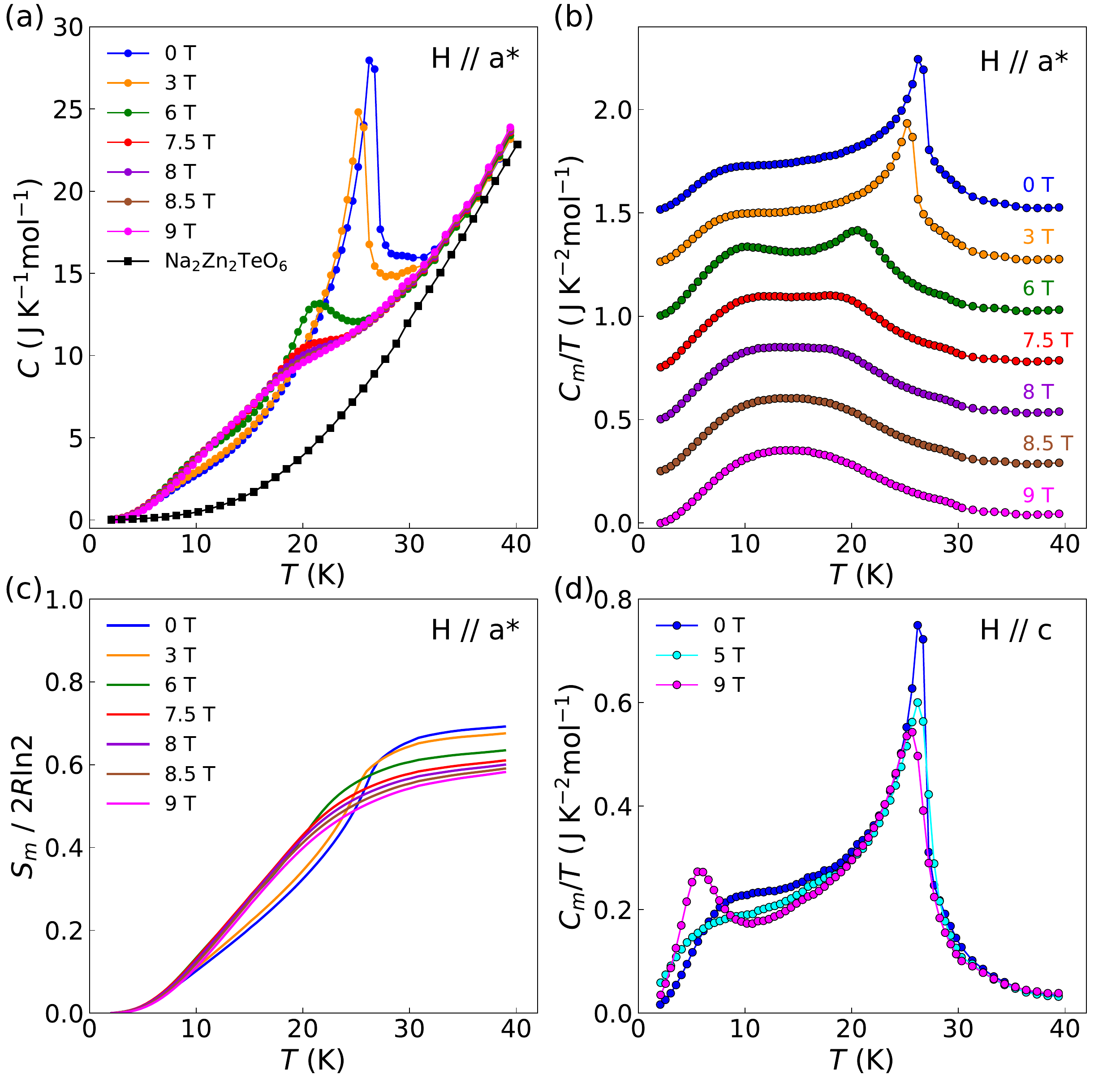}}
	\caption{(a) Specific heat of \ch{Na_2Co_2TeO_6} measured in fields along \textbf{a*}, compared to a nonmagnetic \ch{Na_2Zn_2TeO_6} reference. (b) Magnetic specific heat of \ch{Na_2Co_2TeO_6} divided by $T$, offset in increments of 0.25 J K$^{-2}$ mol$^{-1}$ for clarity. (c) Magnetic entropy released from 2 to 40 K, in units of the full entropy ($2R\ln2$) per unit cell with two pseudospin-1/2. (d) Magnetic specific heat divided by $T$ measured in fields along \textbf{c}.}
	\label{fig3}
\end{figure}

Figure~\ref{fig2}(e) displays isothermal magnetization as functions of fields. The super-linear field dependence, more pronounced for the in-plane fields, is again similar to $\alpha$-\ch{RuCl_3} \cite{BaekPRL2017}. The smaller susceptibility along $\mathbf{c}$ can be explained by $g$-factor anisotropy caused by a nonzero trigonal crystal field on Co$^{2+}$ \cite{AbragamPRSA1951,LinesPR1963}. In line with the large increase of low-$T$ magnetization in Fig.~\ref{fig2}(b) for $H\parallelsum\mathbf{a}^*$ between 5.5 and 6 T, a first-order transition is observed in this field range for $H\parallelsum\mathbf{a}^*$, which gradually decreases with increasing $T$ [Fig.~\ref{fig2}(f)] and disappears above $T_\mathrm{N}$. A similar observation was recently reported and attributed to a spin-flop transition of the zigzag order \cite{XiaoCGD2019}, but the field's in-plane direction was previously unknown. The fact that we observe it with $H\parallelsum\mathbf{a}^*$, \textit{i.e.}, perpendicular to the ordered moments or at 30$^\circ$ from them, but not with $H\parallelsum\mathbf{a}$ which is parallel to the ordered moments in one zigzag domain, is inconsistent with a spin-flop interpretation \cite{BlundellBook}. Instead, the result points towards a field-induced reversal of moment canting perpendicular to the chains, as such canting must exist to give rise to the ferrimagnetism discussed above. Similarly, we believe that the low-$T$ upturn of $M$ in $H\parallelsum\mathbf{c}$ greater than 4 T [Fig.~\ref{fig2}(c)] is related to a change of canting along $\mathbf{c}$.

All magnetic transitions in Fig.~\ref{fig2} leave their signatures in the specific heat. In zero field, the transition into the zigzag phase is signified by a prominent specific-heat peak at $T_\mathrm{N}$, which is rapidly suppressed by $H\parallelsum\mathbf{a}^*$ and becomes no longer noticeable for $H>8$ T [Fig.~\ref{fig3}(a)]. No more peaks are seen in zero field below $T_\mathrm{N}$, consistent with the notion that no transition occurs at the ferrimagnetic compensation point [Fig.~\ref{fig2}(d)], and that the additional susceptibility anomaly seen at 16 K (Fig.~S5 in \cite{SM}) is insignificant. After the peak at $T_\mathrm{N}$ is fully suppressed, a broad hump remains in the $C_m/T$ data [Fig.~\ref{fig3}(b)], where the magnetic specific heat $C_m$ is obtained by subtracting phonon contributions measured on an isostructural \ch{Na_2Zn_2TeO_6} reference crystal \cite{BerthelotJSSC2012,SM}. This broad hump, as in the case of $\alpha$-\ch{RuCl_3} \cite{SearsPRB2017}, may be due to short-range spin correlations. At $H=6$ T, a second broad peak shows up below 10 K [Fig.~\ref{fig3}(b)], which we believe is related to the in-plane canting reversal discussed previously [Fig.~\ref{fig2}(e-f)]; indeed, we find hints for a thermal transition below 10 K related to this also in the 5.5 T and 6 T magnetization data in Fig.~\ref{fig2}(b). The calculated magnetic entropy release from 2 K [$S_m(T)=\int_{2\mathrm{K}}^{T}C_m/TdT$] up to 40 K amounts to only 70\% of the expected molar value $2R\ln2$ (two Co$^{2+}$ per formula unit), and it further decreases with increasing field [Fig.~\ref{fig3}(c)]. This indicates substantial fluctuations on the pseudospin-1/2 degrees of freedom, regardless of the order. The sharp peak at $T_N$ is hardly affected by $H\parallelsum\mathbf{c}$ up to 9 T [Fig.~\ref{fig3}(d)], yet a new peak appears below 10 K in high fields, reminding us of the upturn in $M$ seen at low $T$ in Fig.~\ref{fig2}(c), which we have attributed to canting reversal along $\mathbf{c}$.

\begin{figure}[t]
	\centering{\includegraphics[clip,width=8cm]{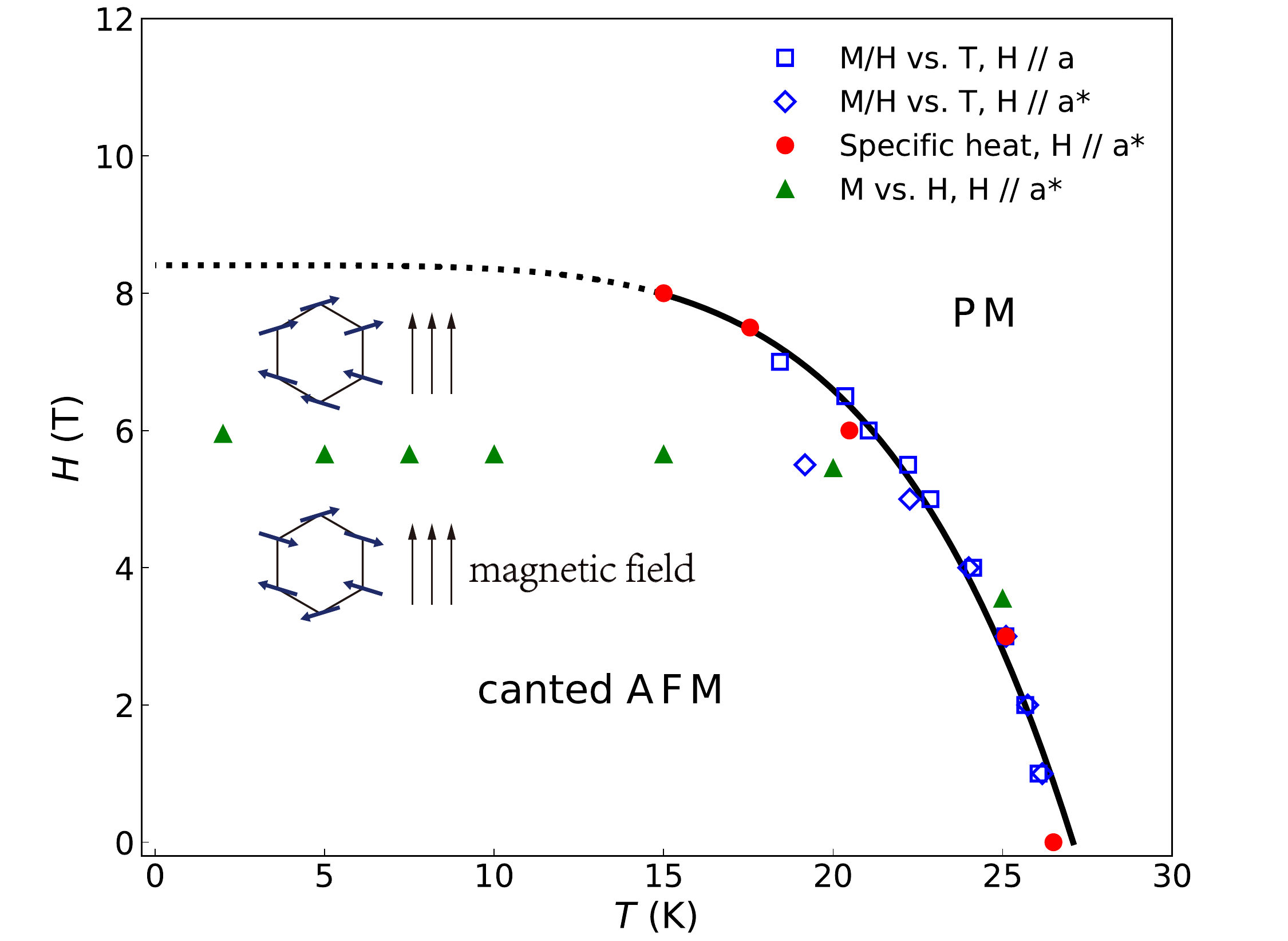}}
	\caption{Phase diagram under in-plane magnetic fields.  The phase boundaries are determined from measurements in Figs.~\ref{fig2} and \ref{fig3}. Solid line is the fitted PM-AFM phase boundary (see text), and dashed line is its extrapolation. Schematic insets illustrate possible in-plane canting reversal associated with the first-order transition observed with $H\parallelsum\mathbf{a}^*$ in Fig.~\ref{fig2}(f).}
	\label{fig4}
\end{figure}

We summarize our results obtained with in-plane fields in Fig.~\ref{fig4}, in which we find a good agreement between phase boundaries determined from different measurements. Our result suggests that the previously identified zigzag phase \cite{LefrancoisPRB2016,BeraPRB2017} is likely a canted AFM phase that features a superposition of dominant zigzag order (moments parallel to the zigzag chains) and minor N\'{e}el order (moments perpendicular to the chains). At low $T$ in $H\gtrsim 6$ T perpendicular to the zigzag chains, the N\'{e}el AFM order can be converted to a FM order through a first-order transition. The paramagnetic-antiferromagnetic (PM-AFM) phase boundary can be fitted with a power-law: $T_\mathrm{N}(H)=17.8(H_\mathrm{c}-H)^n$, with critical field $H_\mathrm{c}= 8.4$ T and critical exponent $n=0.20$. The critical exponent here is rather close to the one in $\alpha$-\ch{RuCl_3} (0.18, \cite{SearsPRB2017}), which implies that the underlying physics might also be similar, in spite of the two systems' rather different crystal and electronic structures.

In summary, we have discovered ferrimagnetism in the AFM phase of \ch{Na_2Co_2TeO_6}, which is naturally explained by a superposition of N\'{e}el-type moment canting on collinear zigzag order. Using magnetic fields transverse to the zigzag chains, we can at least partly reverse the moment canting. Moreover, we show that the AFM order itself is strongly suppressed by in-plane fields, and that the pseudospin-1/2 degrees of freedom strongly fluctuate independent of the order. Some of these results are intriguingly similar to those in the Kitaev magnet $\alpha$-\ch{RuCl_3}, and overall they manifest an intricate phase interplay that is generally expected near boundaries of competing phases, where quantum fluctuations are important. \ch{Na_2Co_2TeO_6} is hence a new and exciting platform for studying physics related to the Kitaev model.

We would like to thank Profs. Gang Chen and Zhengcai Xia for helpful discussions, Prof. Shuang Jia for using their X-ray diffractometer, and Dr. Xintong Li for a careful read of the manuscript. This work is supported by the NSF of China under Grant Nos. 11874069 and 11888101, and by the NBRP of China under Grant Nos. 2018YFA0305602 and 2015CB921302.

\nocite{apsrev42Control}
\bibliographystyle{apsrev4-2}

\bibliography{Reference_NTCO}% Produces the bibliography via BibTeX.

%%%%%%%%%% Merge with supplemental materials %%%%%%%%%%
\pagebreak
\pagebreak
\widetext
\begin{center}
\textbf{\large Supplemental Material for ``Ferrimagnetism and anisotropic phase tunability by magnetic fields in \ch{Na_2Co_2TeO_6}''}
\end{center}
%%%%%%%%%% Merge with supplemental materials %%%%%%%%%%
%%%%%%%%%% Prefix a "S" to all equations, figures, tables and reset the counter %%%%%%%%%%
\setcounter{equation}{0}
\setcounter{figure}{0}
\setcounter{table}{0}
\setcounter{page}{1}
\makeatletter
\renewcommand{\theequation}{S\arabic{equation}}
\renewcommand{\thefigure}{S\arabic{figure}}
\renewcommand{\bibnumfmt}[1]{[S#1]}
\renewcommand{\citenumfont}[1]{S#1}
%%%%%%%%%% Prefix a "S" to all equations, figures, tables and reset the counter %%%%%%%%%%

\section{Additional sample characterizations}

\begin{figure}[H]
	\centering{\includegraphics[clip,width=11cm]{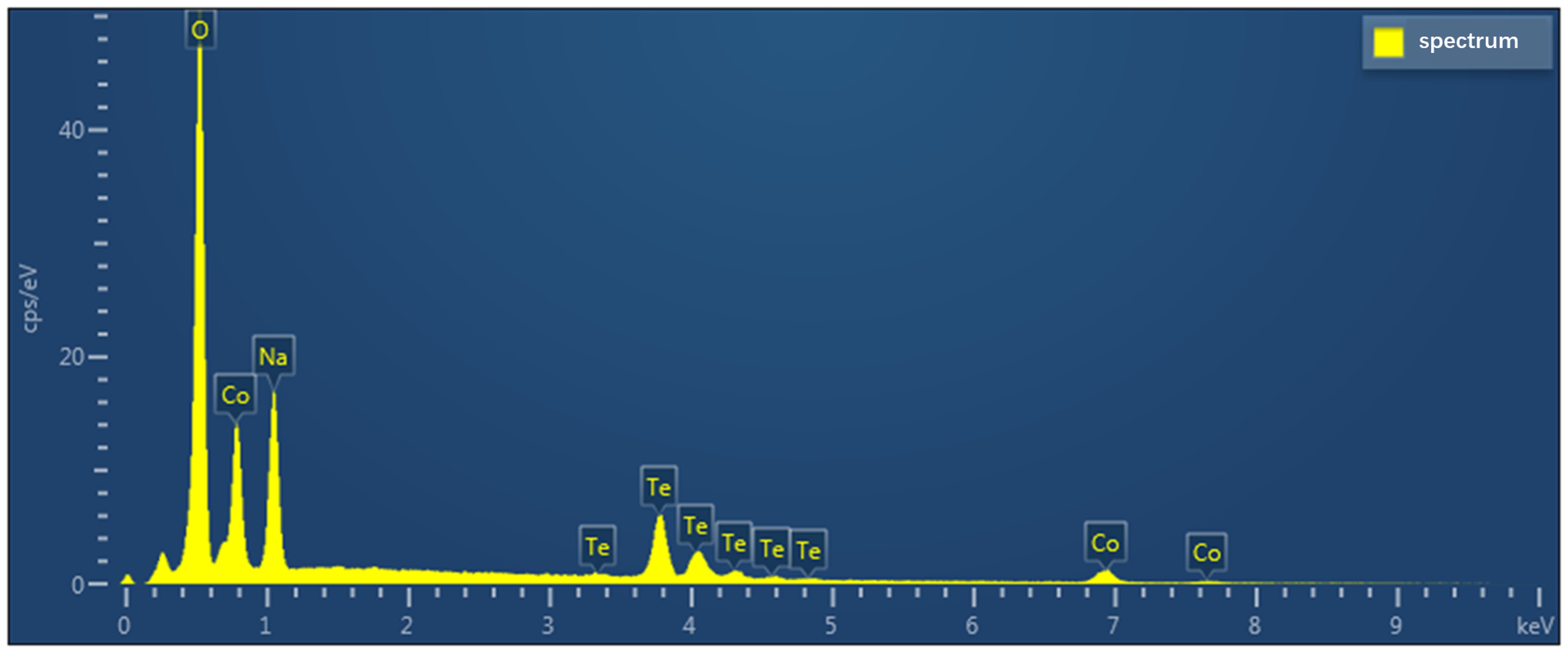}}
	\caption{\label{figs1}Energy dispersive X-ray spectrum at a representative spot on a \ch{Na_2Co_2TeO_6} single crystal.}
\end{figure}

Figure \ref{figs1} shows the energy dispersive X-ray spectrum on a \ch{Na_2Co_2TeO_6} single crystal. The elemental analysis showed in Table \ref{table1} indicates the element content in our sample is in good agreement with the chemical formula.

\begin{figure}[H]
	\centering{\includegraphics[clip,width=8cm]{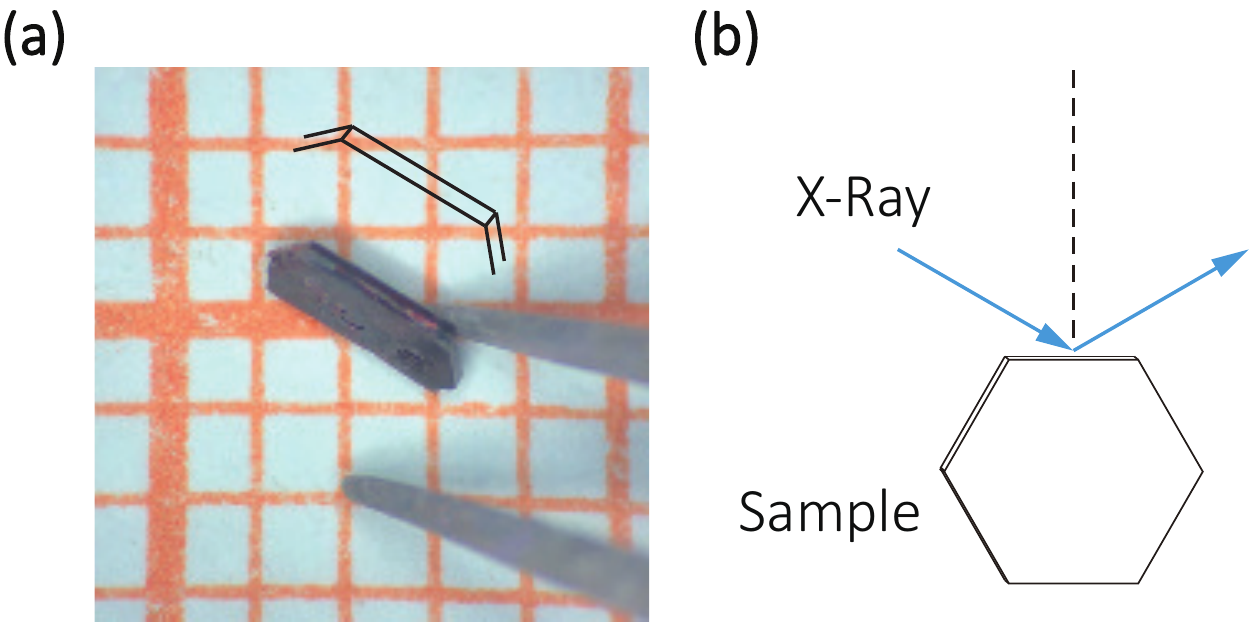}}
	\caption{\label{figs2}(a) The thick \ch{Na_2Co_2TeO_6} single crystal used for X-ray diffraction on $(0, K, 0)$ planes in the main text. (b) Schematic diagram for the XRD measurement.}
\end{figure}

Figure \ref{figs2}(a) shows the thick single crystal used for XRD measurement on $(0, K, 0)$ planes in the main text. In this measurement, we put the ab-plane vertically to make the momentum transfer perpendicular to its natural side, as schematically showed in Fig.~\ref{figs2}(b). The measured intensities and 2$\theta$ values are consistent with $(0, K, 0)$ reflections (see Table \ref{table2}).

\begin{figure}[H]
	\centering
	\includegraphics[width=10cm]{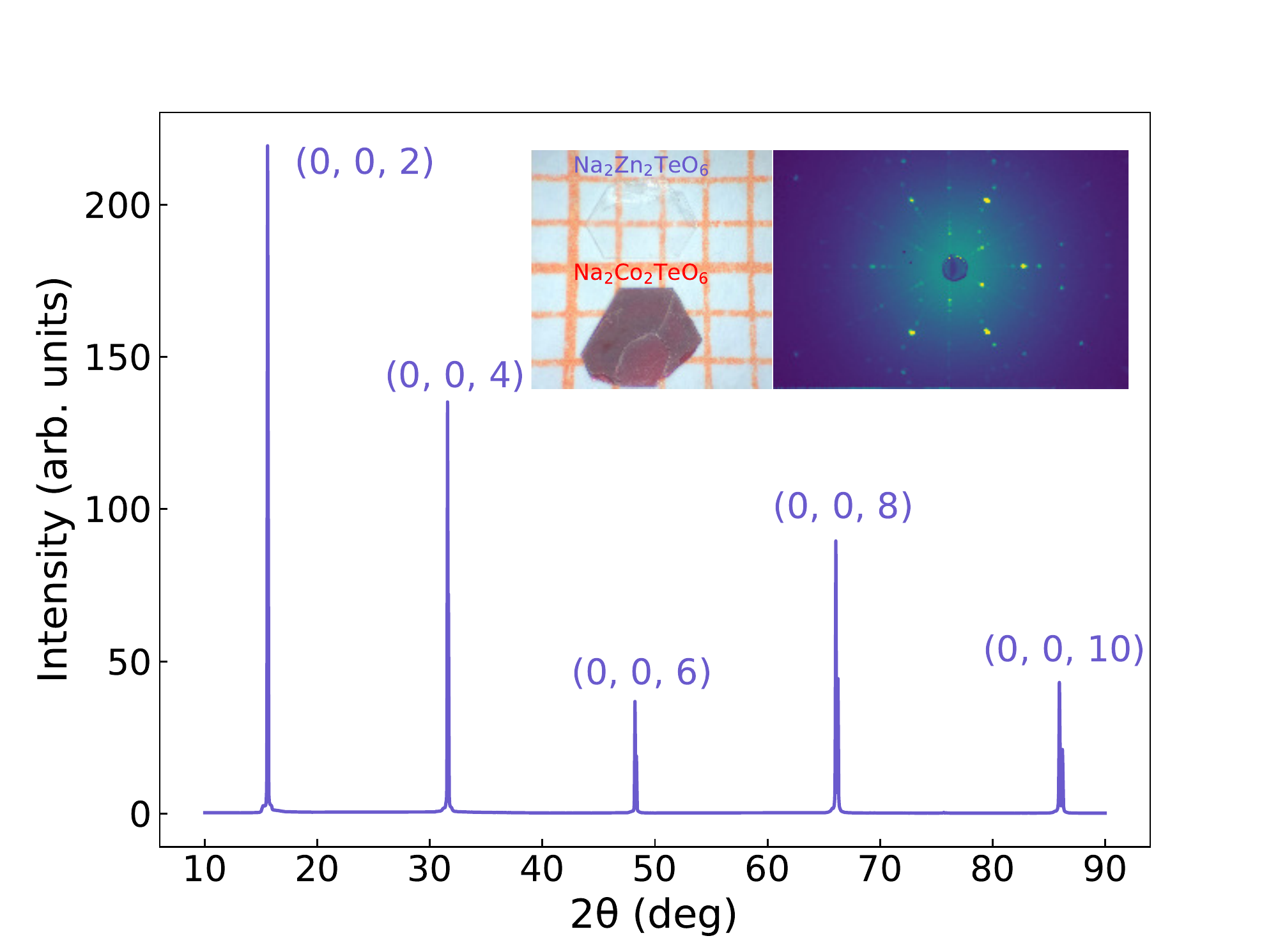}
	\caption{\label{figs3}Single crystal X-ray diffraction pattern in $(0, 0, L)$ planes of \ch{Na_2Zn_2TeO_6}. The inset shows the picture of  \ch{Na_2Co_2TeO_6} and \ch{Na_2Zn_2TeO_6} single crystals (left) and the Laue pattern of \ch{Na_2Zn_2TeO_6} (right).}
\end{figure}

Figure \ref{figs3} shows X-ray diffraction and Laue characterizations of a \ch{Na_2Zn_2TeO_6} single crystal used in our reference specific heat measurement. The results are similar with those for \ch{Na_2Co_2TeO_6}. In the left of the inset, we also present the \ch{Na_2Co_2TeO_6} single crystal used in our magnetization and specific heat experiments.

\begin{figure}[H]	
	\centering
	\includegraphics[width=11cm]{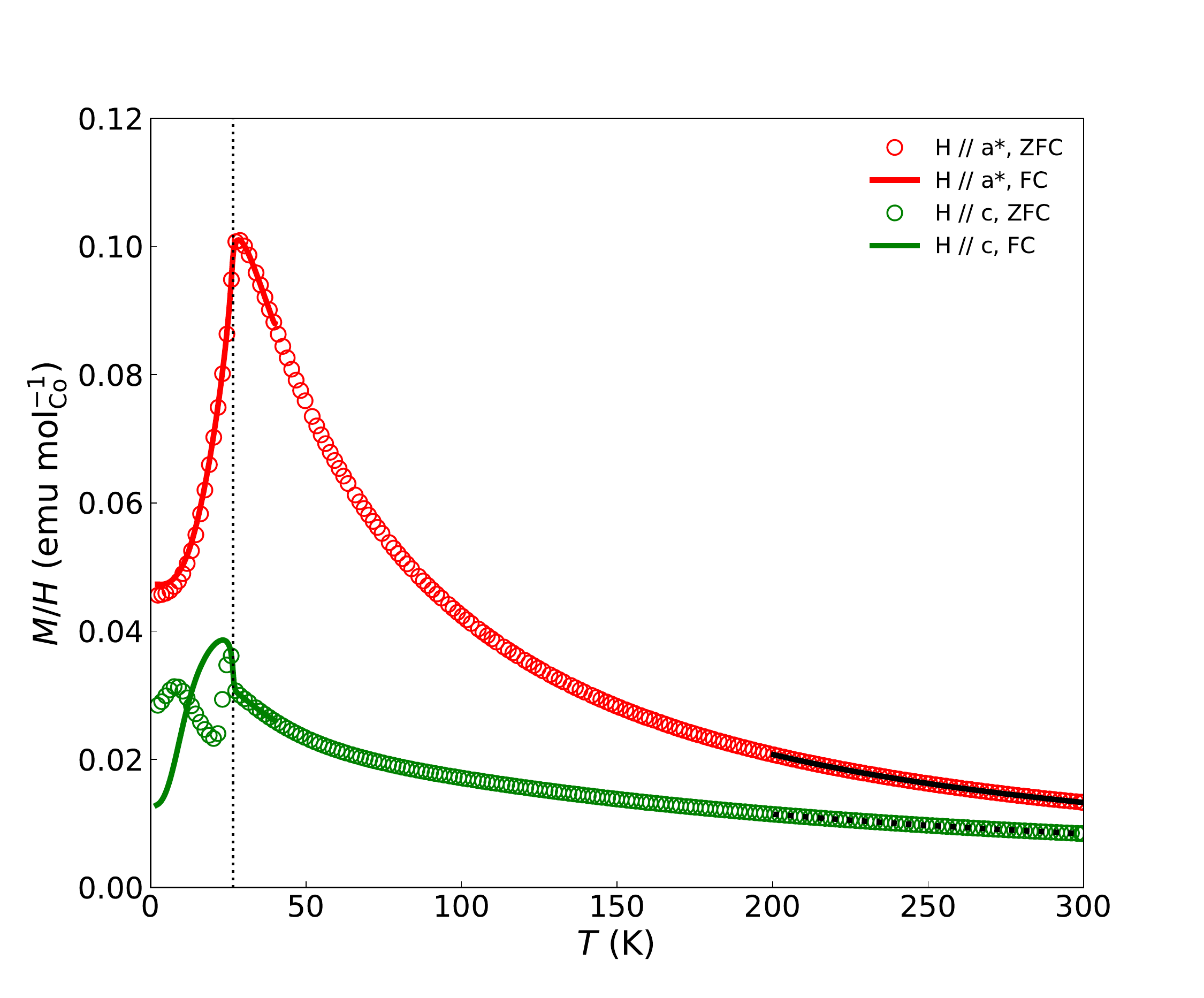}
	\caption{\label{figs4}Temperature dependence of M/H for $H \parallelsum \mathbf{a}^*$ (red) and $H \parallelsum \mathbf{c}$ (green) in ZFC (empty circles) and FC (solid lines) conditions at 0.5 T. Solid and dashed curves in black are the corresponding Curie-Weiss fits. The vertical dashed line indicates $T_N$ = 26.5 K.}
\end{figure}

In Fig.~\ref{figs4} we present magnetization measurements from 2 K to 300 K with $H=0.5$ T. We have made Curie-Weiss fit ($\chi=\chi_0+C/(T-\Theta)$, with $\chi=M/H$) from 200 K to 300 K. The effective magnetic moments are $\mu_{\mathbf{a}*}^{eff}=5.69$ $\mu_B$ and $\mu_{\mathbf{c}}^{eff}=5.31$ $\mu_B$, with the corresponding Weiss temperatures $\Theta_{\mathbf{a}*}=12.5$ K and $\Theta_{\mathbf{c}}=-93.8$ K, respectively. The zero-field-cooling (ZFC) curve for $H \parallelsum \mathbf{c}$ shows some additional kinks below $T_N$. These kinks, as we discussed in the main text, are due to the prominent ferrimagnetic signals along $\mathbf{c}$, which are influenced by small remnant fields during cooling.

\section{Magnetization measurements under small fields}

\begin{figure}[H]	
	\centering{\includegraphics[clip,width=11cm]{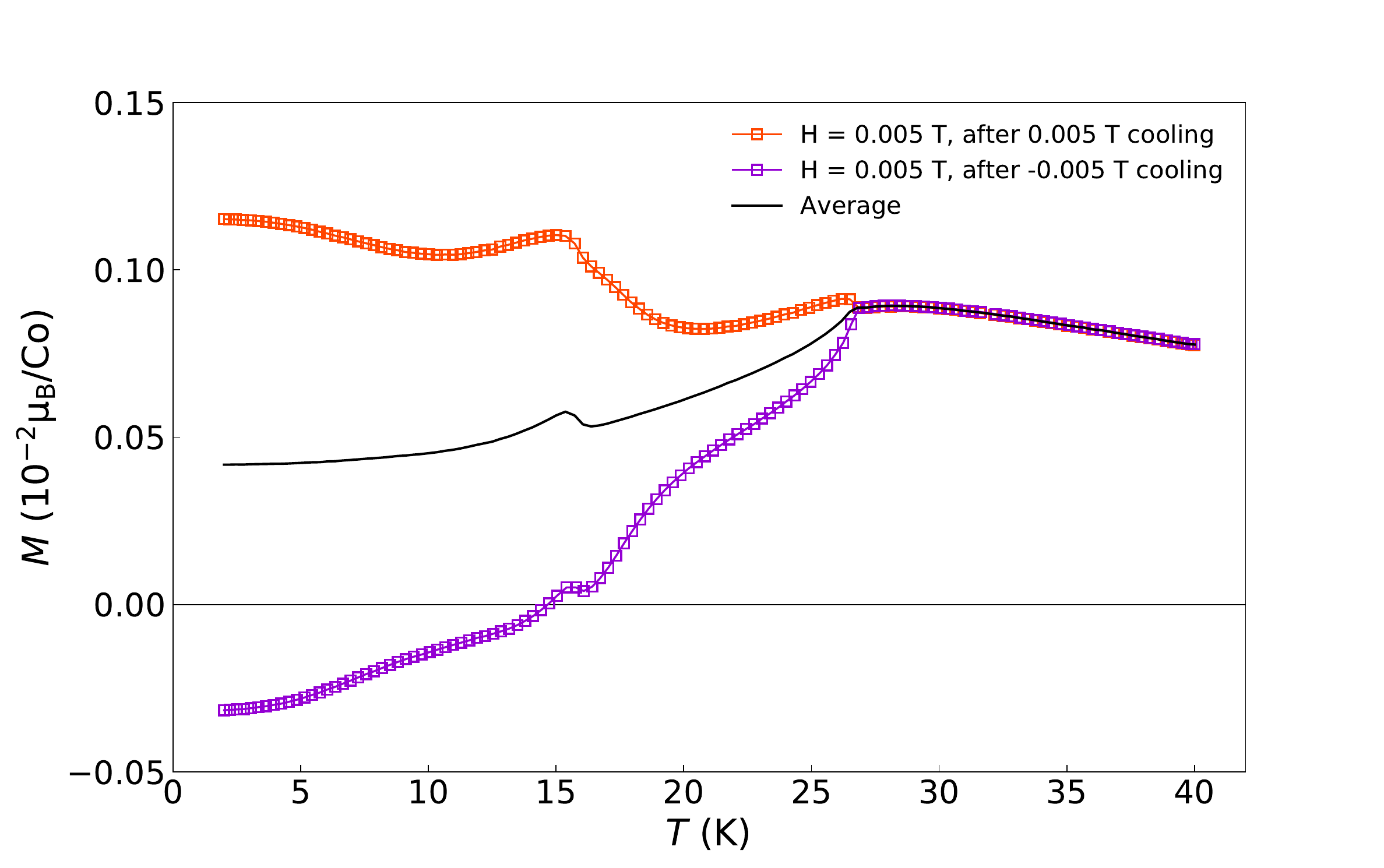}}
	\caption{\label{figs5} Magnetization measured with $H \parallelsum \mathbf{a}$ of 0.005 T, after cooling under 0.005 T and -0.005 T. Black curve indicates the average. The behaviors here have noticeable sample dependence and may be related to disorder and impurity. }
\end{figure}

In Fig.~\ref{figs5}, we present the temperature dependence of the magnetization at 0.005 T after cooling under positive and negative fields (the field used during the measurement was always positive). The slight negative magnetization and the bifurcation below $T_N$ indicate there is a small ferrimagnetic component in this direction. Due to in-plane magnetic domains, this component is expected to have a more complex composition than the $\mathbf{c}$-component. However, since it is much smaller than the latter, we expect it to have even less important effects on the AFM state. Below $T_N$, we can also discern another anomaly at $\sim$16 K,  as observed in previous reports. It remains after making average of the two curves, by which we assume that the ferrimagnetic component is eliminated. This anomaly is therefore most likely different from the ferrimagnetism. % and may be another subordinate phase transition.

\begin{figure}[H]
	\centering
	\includegraphics[clip,width=11cm]{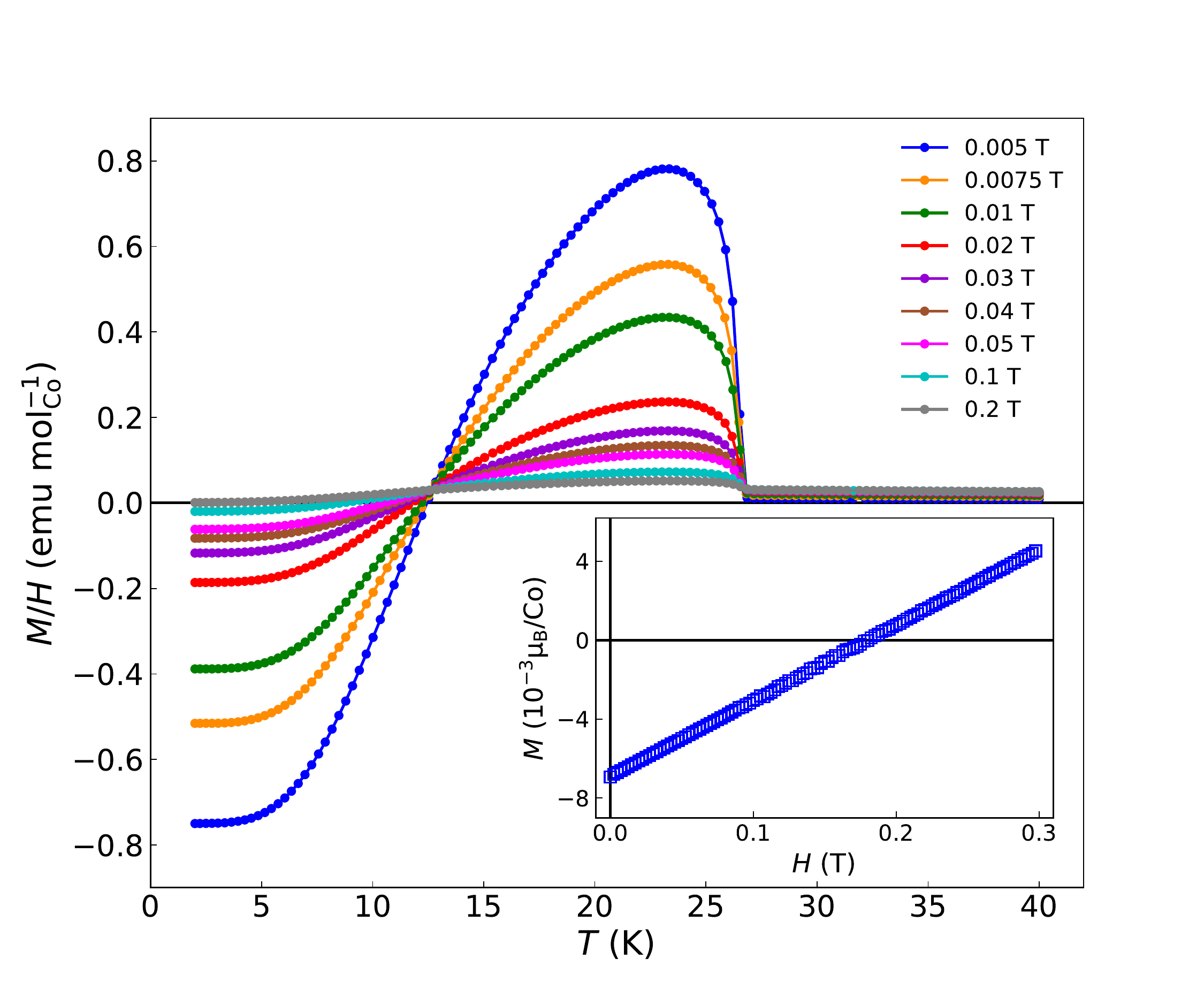}
	\caption{\label{figs6}Temperature dependence of M/H for $H \parallelsum \mathbf{c}$ at various magnetic fields measured after field-cooling. The inset shows isothermal magnetization for $H \parallelsum \mathbf{c}$ at 2 K after cooling under 0.005 T.}
\end{figure}

In Fig.~\ref{figs6} we present the temperature dependence of the magnetization under various magnetic fields along $\mathbf{c}$ in field-cooling (FC) condition. All curves intersect with each other around the compensation point $\sim$12.5 K. In the inset, we present the field dependence of the magnetization at 2 K after cooling under 0.005 T. The magnetization curve has a linear field dependence up to 0.3 T with a negative intercept. This suggests that the ferrimagnetic component has little field dependence. From these observations, we can conclude that the full magnetization consists of a regular, susceptibility-related component and a ferrimagnetic component, the former of which increases approximately linearly with field. The full magnetization then can be expressed as $M(T)=M_\mathrm{ferri}(T)+\chi_c(T)\cdot H$.

\begin{table}[h!]
	\centering
	\newcommand{\tabincell}[2]{\begin{tabular}{@{}#1@{}}#2\end{tabular}}
	\begin{tabular}{|c|c|c|c|}
		\hline
		Element&	\tabincell{c}{X-ray\\line type} & \tabincell{c}{Weight\\percentage}& \tabincell{c}{Atomic\\percentage}\\
		\hline
		Na& K& 12.18 \%& 18.96 \%\\
		\hline
		Co& L& 30.67 \%& 18.63 \%\\
		\hline
		Te& L& 33.45 \%& 9.38 \%\\
		\hline
		O& K& 23.7 \%& 53.03 \%\\
		\hline
		Total&  &  100 \%& 100 \%\\
		\hline
	\end{tabular}
	\caption{Elemental analysis report from the spectrum in Fig.\ref{figs1}.}
	\label{table1}
\end{table}

\begin{table}[H]
	\centering
	\newcommand{\tabincell}[2]{\begin{tabular}{@{}#1@{}}#2\end{tabular}}
	\begin{tabular}{|c|c|c|c|c|}
		\hline
		(H, K, L)&	d ($\rm \AA$)    &$\rm 2\theta$ (deg)    & I/I(0,0,2)    & I/I(0,1,0) \\
		\hline
		(0, 0, 2)& 5.61& 15.79& 1.00& - \\
		\hline
		(0, 0, 4)& 2.80& 31.89& 0.29& - \\
		\hline
		(0, 0, 6)& 1.87& 48.67& 0.04& - \\
		\hline
		(0, 0, 8)& 1.40& 66.66& 0.11& - \\
		\hline
		(0, 0, 10)& 1.12& 86.76& 0.07& - \\
		\hline
		(0, 1, 0)& 4.58& 19.36& - & 1.00 \\
		\hline
		(0, 2, 0)& 2.29& 39.31& - & 0.18\\
		\hline
		(0, 3, 0)& 1.53& 60.60& - & 11.00\\
		\hline
	\end{tabular}
	\caption{Relative X-ray diffraction intensities of selected reflections calculated based on single crystal.}
	\label{table2}
\end{table}

\end{document}